\documentclass[11pt]{article}
\usepackage{amssymb,amsmath,amsfonts}
\usepackage{graphicx}
\usepackage{graphics}
\usepackage{eepic,epsfig}

\@addtoreset{equation}{section}

\makeatother

\begin{document}

\title{Fermionic vacuum polarization by a flat boundary \\
in cosmic string spacetime }
\author{E. R. Bezerra de Mello$^{1}$\thanks{%
E-mail: emello@fisica.ufpb.br}, \thinspace\ A. A. Saharian$^{1,2}$\thanks{%
E-mail: saharian@ysu.am}, \thinspace\ S. V. Abajyan$^{3}$ \vspace{0.3cm} \\
%EndAName
\textit{$^{1}$Departamento de F\'{\i}sica, Universidade Federal da Para\'{\i}%
ba}\\
\textit{58.059-970, Caixa Postal 5.008, Jo\~{a}o Pessoa, PB, Brazil}\vspace{%
0.3cm}\\
\textit{$^2$Department of Physics, Yerevan State University,}\\
\textit{1 Alex Manoogian Street, 0025 Yerevan, Armenia} \vspace{0.3cm}\\
\textit{$^3$Armenian State Pedagogical University,%
}\\
\textit{13 Khandjyan Street, 0010 Yerevan, Armenia}}
\date{}
\maketitle

\begin{abstract}
In this paper we investigate the fermionic condensate and the
renormalized vacuum expectation value (VEV) of the energy-momentum
tensor for a massive fermionic field induced by a flat boundary in
the cosmic string spacetime. In this analysis we assume that the
field operator obeys MIT bag boundary condition on the boundary.
We explicitly decompose the VEVs into the boundary-free and
boundary-induced parts. General formulas are provided for both
parts which are valid for any value of the parameter associated
with the cosmic string. For a massless field, the boundary-free
part in the fermionic condensate and the boundary-induced part in
the energy-momentum tensor vanish. For a massive field the radial
stress is equal to the energy density for both boundary-free and
boundary-induced parts. The boundary-induced part in the stress
along the axis of the cosmic string vanishes. The total energy
density is negative everywhere, whereas the effective pressure
along the azimuthal direction is positive near the boundary and
negative near the cosmic string. We show that for points away from
the boundary, the boundary-induced parts in the fermionic
condensate and in the VEV of the energy-momentum tensor vanish on
the string.
\end{abstract}

\bigskip

PACS numbers: 03.70.+k, 98.80.Cq, 11.27.+d

\bigskip

\section{Introduction}

Non-trivial structure of the vacuum state is among the most profound
predictions of quantum field theory. The properties of the vacuum are
manifested in its response to different external perturbations. The
imposition of boundary conditions on a quantum field is among the simplest
models for the perturbation. The boundary conditions may be induced by real
material media or represent non-trivial topology of the space. In both cases
they modify the structure of the quantum vacuum thus affecting the vacuum
expectation values (VEVs) of physical observables. This phenomenon is known
as the Casimir effect (for reviews see Ref. \cite{Most97}). In the present
paper we consider combined effects from boundaries and non-trivial topology
on the properties of the fermionic vacuum in the cosmic string spacetime.

The cosmic strings are predicted to be formed within the framework of grand
unified theories as a result of symmetry braking phase transitions \cite%
{Vile94}. This type of phase transitions have several cosmological
consequences and provide an important link between particle physics and
cosmology. Although the recent observational data on the cosmic microwave
background radiation have ruled out cosmic strings as the primary source for
primordial density perturbations, they are still candidates for the
generation of a number of interesting physical effects such as the
generation of gravitational waves, gamma ray bursts, high energy cosmic
rays, and enhanced baryon number violation. Recently, cosmic strings
attracted a renewed interest partly because a variant of their formation
mechanism is proposed in the framework of brane inflation \cite{Sara02} (for
a review of the existence of cosmic strings in string theory models see \cite%
{Cope11}).

At distances larger than the core radius, the spacetime geometry for a
straight cosmic string is well approximated by a flat spacetime with a
planar angle deficit. In quantum field theory, the corresponding non-trivial
topology gives rise to the polarization of the quantum vacuum. This problem
has been studied in many papers for scalar, fermionic and electromagnetic
fields \cite{Hell86}-\cite{Site09}. The vacuum polarization by a cosmic
string in various curved backgrounds has been discussed in \cite%
{Davi88,Beze09,Otte10,Beze12a}. Another type of vacuum polarization arises
by the presence of boundaries on which the field operator obeys some
prescribed boundary condition. The analysis of the boundary-induced quantum
effects in the cosmic string spacetime have been developed for scalar \cite%
{Beze06sc,Nest11}, fermionic \cite{Beze08,Beze10}, and electromagnetic
fields \cite{Nest11,Brev95,Beze07} for the geometry of a coaxial cylindrical
boundary. From the point of view of the physics in the exterior region, the
latter can be considered as a simplified model for the nontrivial core of
the cosmic string. The Casimir force for massless scalar fields subject to
Dirichlet and Neumann boundary conditions in the setting of the conical
piston has been recently investigated in \cite{Fucc11}. The Casimir
densities for scalar and electromagnetic fields induced by flat boundaries
perpendicular to the string have been considered in \cite{Beze11,Beze12}.
The topological effects in the generalized cosmic string geometry with a
compact dimension along its axis are investigated in \cite{Beze12b}.
Combined effects of topology and boundaries on the Casimir-Polder force
acting on a polarizable microparticle situated near the cosmic string have
been studied in \cite{Saha11}. In particular, it has been shown that due to
the nontrivial topology this force can be repulsive.

Continuing in this line of investigations, in the present paper we consider
the polarization of the fermionic vacuum by a flat boundary perpendicular to
the string axis on which the field operator is constrained by MIT bag
boundary condition. We evaluate the fermionic condensate (FC) and the VEV of
the energy-momentum tensor. These quantities are among the most important
characteristics of the vacuum state. Although the corresponding operators
are local, they carry an important information about the global properties
of the background spacetime. In addition to describing the physical
structure of the quantum field at a given point, the energy-momentum tensor
acts as the source in Einstein's equations and, therefore, it plays an
important role in modelling a self-consistent dynamics which involves the
gravitational field. The FC is crucial in the models of dynamical chiral
symmetry breaking and in considerations of the stability of fermionic
vacuum. As it will be shown below, in the problem under consideration all
calculations can be performed in a closed form and it constitutes an example
in which the topological and boundary-induced polarizations of the vacuum
can be separated in different contributions. Note that a conical space
appears as an effective background geometry in the long-wavelength
description of certain condensed matter systems such as crystals, liquid
crystals and quantum liquids \cite{Nels02}. The results described in this
paper may shed light on the features of the boundary-induced effects in this
type of medium with a conical defect.

The rest of the paper is organized as follows. In the next section we
describe the geometry of the problem and a complete set of positive- and
negative-energy four-spinors obeying the bag boundary condition on the plate
is constructed. In section \ref{sec:FC} these mode functions are applied for
the evaluation of the FC. Similar calculations for the VEV of the
energy-momentum tensor are presented in section \ref{sec:EMT}. The main
results of the paper are summarized in section \ref{sec:Conc}. The technical
details for the evaluation of the energy density and the azimuthal stress
are described in Appendix \ref{sec:Appendix}.

\section{Geometry of the problem and mode functions}

\label{sec:Modes}

We consider the background spacetime corresponding to an infinitely long
straight cosmic string. Considering the string along the $z$-axis, by using
cylindrical coordinates the line element associated with this spacetime
reads:%
\begin{equation}
ds^{2}=dt^{2}-dr^{2}-r^{2}d\phi ^{2}-dz{}^{2},  \label{ds21}
\end{equation}%
where $0\leqslant \phi \leqslant \phi _{0}=2\pi /q$ and the spatial points $%
(r,\phi ,z)$ and $(r,\phi +\phi _{0},z)$ are to be identified. The line
element (\ref{ds21}) has been derived in \cite{Vile81} in the weak-field
approximation. In this case, the planar angle deficit is related to the mass
per unit length of the string, $\mu _{0}$, by $2\pi -\phi _{0}=8\pi G\mu
_{0} $, where $G$ is the Newton gravitational constant. The validity of the
line element (\ref{ds21}) has been extended beyond linear perturbation
theory by several authors \cite{Gott85}\ (see also \cite{Vile94}). In this
case, the parameter $q$ need not to be close to 1. This is the case also in
condensed matter systems with conical defects. For example, in graphitic
cones one has $q=6/5,3/2,2,3,6$. All this type of cones have been
experimentally observed \cite{Kris97}.

The dynamics of a massive fermionic field in curved spacetime with the
metric tensor $g^{\mu \nu }$ is governed by the Dirac equation
\begin{equation}
i\gamma ^{\mu }\nabla _{\mu }\psi -m\psi =0\ ,  \label{Direq}
\end{equation}%
with the covariant derivative operator
\begin{equation}
\nabla _{\mu }=\partial _{\mu }+\Gamma _{\mu }\ , \quad \Gamma
_{\mu }=\gamma ^{\nu }\gamma _{\nu ;\mu }/4 \ . \label{Covder}
\end{equation}%
Here $\gamma ^{\mu }$ are the Dirac matrices in curved spacetime
and $\Gamma _{\mu }$ is the spin connection. Here and below, $\mu
=0,1,2,3$ correspond to the coordinates $t,r,\phi ,z$,
respectively, and the semicolon means the standard covariant
derivative for vector fields.

In the present paper we are interested in quantum vacuum effects induced by
a flat boundary located at $z=0$. We assume that on the boundary the field
operator obeys MIT bag boundary condition:
\begin{equation}
\left( 1+i\gamma ^{\mu }n_{\mu }\right) \psi =0\ ,\quad z=0,  \label{BagCond}
\end{equation}%
where $n_{\mu }$ is the outward-pointing normal to the boundary. In the
discussion below we will consider the region $z>0$ with $n_{\mu }=(0,0,0,-1)$%
. For the evaluation of the VEVs we will use the direct mode summation
procedure. In this approach one needs to have the complete set of the
eigenfunctions satisfying boundary condition (\ref{BagCond}).

In the discussion below the gamma matrices will be taken in the Dirac
representation:%
\begin{equation}
\gamma ^{0}=\left(
\begin{array}{cc}
1 & 0 \\
0 & -1%
\end{array}%
\right) ,\;\gamma ^{l}=\left(
\begin{array}{cc}
0 & \beta ^{l} \\
-\beta ^{l} & 0%
\end{array}%
\right) ,  \label{gam0l}
\end{equation}%
with $\beta ^{l}$, $l=1,2,3$, being the Pauli matrices in cylindrical
coordinates with a planar angle deficit:
\begin{equation*}
\beta ^{1}=\left(
\begin{array}{cc}
0 & e^{-iq\phi } \\
e^{iq\phi } & 0%
\end{array}%
\right) ,\;\beta ^{2}=-\frac{i}{r}\left(
\begin{array}{cc}
0 & e^{-iq\phi } \\
-e^{iq\phi } & 0%
\end{array}%
\right) ,
\end{equation*}%
and $\beta ^{3}=\mathrm{diag}(1,-1)$. We have the standard anticommutation
relation $\beta ^{l}\beta ^{n}+\beta ^{n}\beta ^{l}=-2g^{ln}$. For the spin
connection one gets%
\begin{equation}
\Gamma _{\mu }=\frac{1-q}{2}r\gamma ^{1}\gamma ^{2}\delta _{\mu }^{2}.
\label{gammu}
\end{equation}%
By taking into account that $\gamma ^{\mu }\Gamma _{\mu }=(1-q)\gamma
^{1}/(2r)$, the Dirac equation takes the form%
\begin{equation}
\gamma ^{\mu }\partial _{\mu }\psi +\frac{1-q}{2r}\gamma ^{1}\psi +im\psi =0.
\label{Direq1}
\end{equation}%
First we consider the positive-energy solutions of this equation obeying the
boundary condition (\ref{BagCond}).

\subsection{Positive-energy modes}

For the positive-energy solutions the time dependence has the form $%
e^{-i\omega t}$. Decomposing the four-spinor $\psi $ into two-component
spinors,
\begin{equation}
\psi =\left(
\begin{array}{c}
\varphi \\
\chi%
\end{array}%
\right) ,  \label{psidec}
\end{equation}%
from Eq. (\ref{Direq1}) for the upper and lower components we find the set
of equations
\begin{eqnarray}
\left( \beta ^{l}\partial _{l}\mathbf{+}\frac{1-q}{2r}\beta ^{1}\right)
\varphi -i\left( \omega +m\right) \chi &=&0,  \notag \\
\left( \beta ^{l}\partial _{l}\mathbf{+}\frac{1-q}{2r}\beta ^{1}\right) \chi
-i\left( \omega -m\right) \varphi &=&0,  \label{phieq}
\end{eqnarray}%
From (\ref{phieq}) the following equation is obtained for the upper
component:%
\begin{equation}
\left[ -g^{ij}\partial _{i}\partial _{j}\mathbf{+}\frac{1}{r}\partial _{1}+%
\frac{q-1}{r}\beta ^{1}\beta ^{2}\partial _{2}\mathbf{-}\frac{(q-1)^{2}}{%
4r^{2}}+\omega ^{2}-m^{2}\right] \varphi =0,  \label{Equpper}
\end{equation}%
where $i,j=1,2,3$.

Further decomposing the two-spinor $\varphi $ as
\begin{equation}
\varphi =\left(
\begin{array}{c}
\varphi _{1} \\
\varphi _{2}%
\end{array}%
\right) ,  \label{phidec}
\end{equation}%
we can present the solutions for the separate components in the form%
\begin{equation}
\varphi _{l}=f_{l}(r)\left( C_{\varphi 1}^{(l)}e^{ikz}+C_{\varphi
2}^{(l)}e^{-ikz}\right) e^{i\left( qn_{l}\phi -\omega t\right) },\;l=1,2,
\label{phil}
\end{equation}%
with $0\leqslant k<\infty $ and $n_{l}=0,\pm 1,\pm 2,\ldots $. Substituting (%
\ref{phil}) into (\ref{Equpper}), we can see that for the radial functions
the Bessel equation is obtained with the solutions regular at the string $%
f_{l}(r)=J_{\beta _{l}}(\lambda r)$, where%
\begin{equation}
\beta _{l}=|qn_{l}-(-1)^{l}(q-1)/2|,  \label{betl}
\end{equation}%
and $\lambda =\sqrt{\omega ^{2}-k^{2}-m^{2}}$. By using (\ref{phil}), for
the components of the lower two-spinor,
\begin{equation}
\chi =\left(
\begin{array}{c}
\chi _{1} \\
\chi _{2}%
\end{array}%
\right) ,  \label{xidec}
\end{equation}%
we get%
\begin{equation}
\chi _{l}=J_{\beta _{l}}(\lambda r)\left( B_{\varphi
1}^{(l)}e^{ikz}+B_{\varphi 2}^{(l)}e^{-ikz}\right) e^{i\left( qn_{l}\phi
-\omega t\right) },  \label{xil}
\end{equation}%
with the relations $n_{2}=n_{1}+1$, $\beta _{2}=\beta _{1}+\epsilon _{n_{1}}$%
, and we have defined
\begin{equation}
\epsilon _{n_{1}}=\left\{
\begin{array}{cc}
1, & n_{1}\geqslant 0 \\
-1, & n_{1}<0%
\end{array}%
\right. .  \label{epsn1}
\end{equation}%
The coefficients in (\ref{xil}) are defined by the expressions
\begin{eqnarray}
B_{\varphi 1}^{(1)} &=&\frac{kC_{\varphi 1}^{(1)}-i\lambda \epsilon
_{n_{1}}C_{\varphi 1}^{(2)}}{\omega +m},\;B_{\varphi 2}^{(1)}=-\frac{%
kC_{\varphi 2}^{(1)}+i\lambda \epsilon _{n_{1}}C_{\varphi 2}^{(2)}}{\omega +m%
},  \notag \\
B_{\varphi 1}^{(2)} &=&-\frac{kC_{\varphi 1}^{(2)}-i\lambda \epsilon
_{n_{1}}C_{\varphi 1}^{(1)}}{\omega +m},\;B_{\varphi 2}^{(2)}=\frac{%
kC_{\varphi 2}^{(2)}+i\lambda \epsilon _{n_{1}}C_{\varphi 2}^{(1)}}{\omega +m%
}.  \label{BC}
\end{eqnarray}

Hence, the positive-energy solution to the Dirac equation has the form ($%
n_{1}=n$)%
\begin{equation}
\psi _{\sigma }^{(+)}=\left(
\begin{array}{c}
\left( C_{\varphi 1}^{(1)}e^{ikz}+C_{\varphi 2}^{(1)}e^{-ikz}\right)
J_{\beta }(\lambda r) \\
\left( C_{\varphi 1}^{(2)}e^{ikz}+C_{\varphi 2}^{(2)}e^{-ikz}\right)
J_{\beta +\epsilon _{n}}(\lambda r)e^{iq\phi } \\
\left( B_{\varphi 1}^{(1)}e^{ikz}+B_{\varphi 2}^{(1)}e^{-ikz}\right)
J_{\beta }(\lambda r) \\
\left( B_{\varphi 1}^{(2)}e^{ikz}+B_{\varphi 2}^{(2)}e^{-ikz}\right)
J_{\beta +\epsilon _{n}}(\lambda r)e^{iq\phi }%
\end{array}%
\right) e^{i\left( qn\phi -\omega t\right) },  \label{psi+}
\end{equation}%
where%
\begin{equation}
\beta =|q(n+1/2)-1/2|,  \label{betl1}
\end{equation}%
and $\sigma $ is a collective notation for the set of quantum numbers
specifying the solution (see below). We can see that the four-spinor $\psi
_{\sigma }^{(+)}$ is an eigenfunction of the projection of the total
momentum along the cosmic string:%
\begin{equation}
\widehat{J}_{3}\psi _{\sigma }^{(+)}=\left( -i\partial _{\phi }+i\frac{q}{2}%
r\gamma ^{1}\gamma ^{2}\right) \psi _{\sigma }^{(+)}=qj\psi _{\sigma }^{(+)},
\label{J3psi}
\end{equation}%
with
\begin{equation}
j=n+1/2.  \label{j}
\end{equation}%
Note that $\epsilon _{j}=\epsilon _{n}$.

Now we impose the boundary condition (\ref{BagCond}) on the four-spinor (\ref%
{psi+}). From this condition the following two equations are obtained for
the upper and lower components:%
\begin{equation}
\varphi _{1}-i\chi _{1}=0,\;\varphi _{2}+i\chi _{2}=0\;\text{at}\;z=0.
\label{BCphixi}
\end{equation}%
By using these relations, the coefficients in (\ref{psi+}) are expressed in
terms of $C_{\varphi 1}^{(1)}$ and $C_{\varphi 2}^{(1)}$ as
\begin{eqnarray}
C_{\varphi 1}^{(2)} &=&\frac{\epsilon _{n}}{ik\lambda }\left\{ \left[
k^{2}+m\left( \omega +m\right) \right] C_{\varphi 1}^{(1)}+\left( \omega
+m\right) \left( m+ik\right) C_{\varphi 2}^{(1)}\right\} ,  \notag \\
C_{\varphi 2}^{(2)} &=&-\frac{\epsilon _{n}}{ik\lambda }\left\{ \left(
\omega +m\right) \left( m-ik\right) C_{\varphi 1}^{(1)}+\left[ k^{2}+m\left(
\omega +m\right) \right] C_{\varphi 2}^{(1)}\right\} ,  \label{C21}
\end{eqnarray}%
and%
\begin{eqnarray}
B_{\varphi 1}^{(1)} &=&-\frac{m}{k}C_{\varphi 1}^{(1)}-\left( \frac{m}{k}%
+i\right) C_{\varphi 2}^{(1)},  \notag \\
B_{\varphi 2}^{(1)} &=&\left( \frac{m}{k}-i\right) C_{\varphi 1}^{(1)}+\frac{%
m}{k}C_{\varphi 2}^{(1)},  \notag \\
B_{\varphi 1}^{(2)} &=&i\frac{\epsilon _{n}}{\lambda }\left[ \omega
C_{\varphi 1}^{(1)}+\left( m+ik\right) C_{\varphi 2}^{(1)}\right] ,  \notag
\\
B_{\varphi 2}^{(2)} &=&i\frac{\epsilon _{n}}{\lambda }\left[ \left(
m-ik\right) C_{\varphi 1}^{(1)}+\omega C_{\varphi 2}^{(1)}\right] .
\label{B11}
\end{eqnarray}%
One of the coefficients $C_{\varphi 1}^{(1)}$ and $C_{\varphi 2}^{(1)}$ is
determined from the normalization condition obeyed by the wave function and,
hence, one coefficient remains arbitrary. In order to determine the latter
some additional condition should be imposed. A similar situation to find
fermionic wave function was present for the problem with cylindrical
boundary \cite{Beze08}. However, as it can be seen, the additional condition
used there is not appropriate for the present case.

In order to specify the second constant we impose the condition%
\begin{equation}
B_{\varphi 2}^{(2)}/B_{\varphi 1}^{(2)}=-C_{\varphi 2}^{(1)}/C_{\varphi
1}^{(1)}.  \label{AdCond}
\end{equation}%
By taking into account Eq. (\ref{B11}), this leads to the following relation%
\begin{equation}
C_{\varphi 2}^{(1)}=i\kappa _{s}C_{\varphi 1}^{(1)},  \label{relC}
\end{equation}%
where we have defined
\begin{equation}
\kappa _{s}=\frac{\omega +s\lambda }{k-im},\;s=\pm 1.  \label{kapas}
\end{equation}%
Note that we have the relation $\kappa _{s}\kappa _{-s}^{\ast }=1$.

With the relation (\ref{relC}) all coefficients are expressed in terms of $%
C_{\varphi 1}^{(1)}$. Introducing the notations%
\begin{eqnarray}
f_{\pm }(z) &=&e^{ikz}\pm i\kappa _{s}e^{-ikz},  \notag \\
g_{\pm }(z) &=&\left( \omega \pm m\right) f_{\pm }(z)+s\lambda f_{\mp }(z),
\label{fg}
\end{eqnarray}%
the positive-energy solutions are written in the form%
\begin{equation}
\psi _{\sigma }^{(+)}=C_{\varphi 1}^{(1)}\left(
\begin{array}{c}
f_{+}(z)J_{\beta }(\lambda r)e^{-iq\phi /2} \\
i\frac{\epsilon _{j}s}{k}g_{+}(z)J_{\beta +\epsilon _{j}}(\lambda
r)e^{iq\phi /2} \\
\frac{1}{k}g_{-}(z)J_{\beta }(\lambda r)e^{-iq\phi /2} \\
-i\epsilon _{j}sf_{-}(z)J_{\beta +\epsilon _{j}}(\lambda r)e^{iq\phi /2}%
\end{array}%
\right) e^{i\left( qj\phi -\omega t\right) },  \label{psi+1}
\end{equation}%
where $j=\pm 1/2,\pm 3/2,\ldots $, and
\begin{equation}
\beta =\beta _{1}=q|j|-\epsilon _{j}/2.  \label{beta}
\end{equation}%
Now the set of quantum numbers is specified to $\sigma =(\lambda ,k,j,s)$
and
\begin{equation}
\omega =\sqrt{\lambda ^{2}+k^{2}+m^{2}}.  \label{omega}
\end{equation}

It can be checked that the mode functions (\ref{psi+1}) are orthogonal: $%
\int d^{3}x\sqrt{|g|}\psi _{\sigma }^{(+)+}\psi _{\sigma ^{\prime }}^{(+)}=0$
for $\sigma \neq \sigma ^{\prime }$. The remained coefficient $C_{\varphi
1}^{(1)}$ is found from the normalization condition%
\begin{equation}
\int d^{3}x\sqrt{|g|}\psi _{\sigma }^{(+)+}\psi _{\sigma ^{\prime
}}^{(+)}=\delta (\lambda ^{\prime }-\lambda )\delta (k^{\prime }-k)\delta
_{j^{\prime }j}\delta _{s^{\prime }s}\ .  \label{Norm}
\end{equation}%
By taking into account that
\begin{equation}
\int_{0}^{\infty }drrJ_{\beta }(\lambda r)J_{\beta }(\lambda ^{\prime }r)=%
\frac{1}{\lambda }\delta (\lambda -\lambda ^{\prime }),  \label{BessInt}
\end{equation}%
we find%
\begin{equation}
|C_{\varphi 1}^{(1)}|^{2}=\frac{k^{2}\lambda }{8\pi \phi _{0}\omega (\omega
+s\lambda )}.  \label{Cphi}
\end{equation}%
It can be seen that, the positive-energy mode functions (\ref{psi+1}), with
the normalization coefficient given by (\ref{Cphi}), are invariant, up to a
phase, with respect to the transformation $(k,s)\rightarrow (-k,-s)$.

\subsection{Negative-energy modes}

The time-dependence of the negative-energy mode function is in the form $%
e^{i\omega t}$. For the lower two-spinor, $\chi $, we obtain the same
equation as in (\ref{Equpper}). For the corresponding upper and lower
components one finds%
\begin{equation}
\chi _{l}=J_{\beta _{l}}(\lambda r)(C_{\chi 1}^{(l)}e^{ikz}+C_{\chi
2}^{(l)}e^{-ikz})e^{i\left( qn_{l}\phi +\omega t\right) },\;l=1,2,
\label{xilneg}
\end{equation}%
with the notations (\ref{betl}). Substituting these expressions into (\ref%
{phieq}), for the upper components we find the expressions%
\begin{equation}
\varphi _{l}=(B_{\chi 1}^{(l)}e^{ikz}+B_{\chi 2}^{(l)}e^{-ikz})J_{\beta
_{l}}(\lambda r)e^{i\left( qn_{l}\phi +\omega t\right) },  \label{phi12neg}
\end{equation}%
and the relations $n_{2}=n_{1}+1$, $\beta _{2}=\beta _{1}+\epsilon _{n_{1}}$%
. Denoting $n_{1}=n$, the coefficients in (\ref{phi12neg}) are given by the
expressions%
\begin{eqnarray}
B_{\chi 1}^{(1)} &=&-\frac{kC_{\chi 1}^{(1)}-i\lambda \epsilon _{n}C_{\chi
1}^{(2)}}{\omega +m},\;B_{\chi 2}^{(1)}=\frac{kC_{\chi 2}^{(1)}+i\lambda
\epsilon _{n}C_{\chi 2}^{(2)}}{\omega +m},  \notag \\
B_{\chi 1}^{(2)} &=&\frac{kC_{\chi 1}^{(2)}-i\lambda \epsilon _{n}C_{\chi
\varphi 1}^{(1)}}{\omega +m},\;B_{\chi 2}^{(2)}=-\frac{kC_{\chi
2}^{(2)}+i\lambda \epsilon _{n}C_{\chi \varphi 2}^{(1)}}{\omega +m}.
\label{Bxi}
\end{eqnarray}%
As a result, he negative energy modes are presented as%
\begin{equation}
\psi _{\sigma }^{(-)}=\left(
\begin{array}{c}
(B_{\chi 1}^{(1)}e^{ikz}+B_{\chi 2}^{(1)}e^{-ikz})J_{\beta }(\lambda r) \\
(B_{\chi 1}^{(2)}e^{ikz}+B_{\chi 2}^{(2)}e^{-ikz})J_{\beta +\epsilon
_{n}}(\lambda r)e^{iq\phi } \\
(C_{\chi 1}^{(1)}e^{ikz}+C_{\chi 2}^{(1)}e^{-ikz})J_{\beta }(\lambda r) \\
(C_{\chi 1}^{(2)}e^{ikz}+C_{\chi 2}^{(2)}e^{-ikz})J_{\beta +\epsilon
_{n}}(\lambda r)e^{iq\phi }%
\end{array}%
\right) e^{i\left( qn\phi +\omega t\right) },  \label{psi-}
\end{equation}%
with $\beta $ defined in (\ref{betl1}).

From the boundary conditions (\ref{BCphixi}) the coefficients are expressed
in terms of $C_{\chi 1}^{(1)}$ and $C_{\chi 2}^{(1)}$:
\begin{eqnarray}
C_{\chi 1}^{(2)} &=&\frac{\epsilon _{n}}{ik\lambda }\left\{ \left[
k^{2}+m\left( \omega +m\right) \right] C_{\chi 1}^{(1)}+\left( \omega
+m\right) \left( m+ik\right) C_{\chi 2}^{(1)}\right\} ,  \notag \\
C_{\chi 2}^{(2)} &=&-\frac{\epsilon _{n}}{ik\lambda }\left\{ \left( \omega
+m\right) \left( m-ik\right) C_{\chi 1}^{(1)}+\left[ k^{2}+m\left( \omega
+m\right) \right] C_{\chi 2}^{(1)}\right\} ,  \label{Cxi}
\end{eqnarray}%
and%
\begin{eqnarray}
B_{\chi 1}^{(1)} &=&\frac{m}{k}C_{\chi 1}^{(1)}+\left( \frac{m}{k}+i\right)
C_{\chi 2}^{(1)},  \notag \\
B_{\chi 2}^{(1)} &=&-\left( \frac{m}{k}-i\right) C_{\chi 1}^{(1)}-\frac{m}{k}%
C_{\chi 2}^{(1)},  \notag \\
B_{\chi 1}^{(2)} &=&\frac{\epsilon _{n}}{i\lambda }\left[ \omega C_{\chi
1}^{(1)}+\left( m+ik\right) C_{\chi 2}^{(1)}\right] ,  \notag \\
B_{\chi 2}^{(2)} &=&\frac{\epsilon _{n}}{i\lambda }\left[ \left( m-ik\right)
C_{\chi 1}^{(1)}+\omega C_{\chi 2}^{(1)}\right] .  \label{Bxi2}
\end{eqnarray}%
Similar to the case of positive-energy modes, an additional condition should
be imposed. As such a condition we impose%
\begin{equation}
B_{\chi 2}^{(2)}/B_{\chi 1}^{(2)}=-C_{\chi 2}^{(1)}/C_{\chi 1}^{(1)}.
\label{AdCondNeg}
\end{equation}%
Taking into account (\ref{Bxi2}), this gives%
\begin{equation}
C_{\chi 2}^{(1)}=i\kappa _{s}C_{\chi 1}^{(1)},\;s=\pm 1,  \label{AdCondNeg2}
\end{equation}%
with $\kappa _{s}$ defined in (\ref{kapas}).

As a result, the negative-energy mode functions are written in the form%
\begin{equation}
\psi _{\sigma }^{(-)}=C_{\chi 1}^{(1)}\left(
\begin{array}{c}
-\frac{1}{k}g_{-}(z)J_{\beta }(\lambda r)e^{-iq\phi /2} \\
i\epsilon _{j}sf_{-}(z)J_{\beta +\epsilon _{j}}(\lambda r)e^{iq\phi /2} \\
f_{+}(z)J_{\beta }(\lambda r)e^{-iq\phi /2} \\
i\frac{\epsilon _{j}s}{k}g_{+}(z)J_{\beta +\epsilon _{j}}(\lambda
r)e^{iq\phi /2}%
\end{array}%
\right) e^{i\left( qj\phi +\omega t\right) },  \label{psi-1}
\end{equation}%
with the same notations as in (\ref{psi+1}). The function is an
eigenfunction for the projection of the total angular momentum on the $z$%
-axis, $\widehat{J}_{3}\psi _{\sigma }^{(-)}=qj\psi _{\sigma }^{(-)}$, and
we have introduced the related quantum number $j=n+1/2$. It can be checked
that the modes (\ref{psi-1}) obey the orthogonality relations $\int d^{3}x%
\sqrt{|g|}\psi _{\sigma ^{\prime }}^{(\pm )+}\psi _{\sigma }^{(-)}=0$ for $%
\sigma \neq \sigma ^{\prime }$. From the normalization condition
we can see that $|C_{\chi 1}^{(1)}|^{2}=|C_{\varphi
1}^{(1)}|^{2}$. Note that we have a correspondence between the
negative- and positive-energy modes given by the relation (up to a
phase)
\begin{equation}
\psi _{\sigma }^{(-)}e^{-i\omega t}=\gamma
^{5}\gamma ^{0}\psi _{\sigma }^{(+)}e^{i\omega t}, \label{psirel}
\end{equation}
where in cylindrical coordinates $\gamma ^{5}=ir\gamma ^{0}\gamma
^{1}\gamma ^{2}\gamma ^{3}$. This relation corresponds to the PCT
transformation of four-spinors.

The wave functions obtained in this section can be used for the
investigation of various quantum electrodynamics effects around
the cosmic string in the presence of boundary involving electrons
and positrons (for example, bremsstrahlung and electron-positron
pair production by a photon). In what follows we use these
functions for the evaluation of the FC and the VEV of the
energy-momentum tensor.

\section{Fermionic condensate}

\label{sec:FC}

Having the complete set of mode functions for the fermionic field, we can
evaluate the effects of the boundary on the FC, $\langle 0|\bar{\psi}\psi
|0\rangle $, where $|0\rangle $ is the vacuum state in the geometry of a
cosmic string with flat boundary and $\bar{\psi}=\psi ^{+}\gamma ^{0}$ is
the Dirac adjoint. Expanding the field operator in terms of the complete set
$\{\psi _{\sigma }^{(-)},\psi _{\sigma }^{(+)}\}$ and by using the
definition of the vacuum state, the following formula is obtained for the
condensate:
\begin{equation}
\langle 0|\bar{\psi}\psi |0\rangle =\sum_{\sigma }\bar{\psi}_{\sigma
}^{(-)}\psi _{\sigma }^{(-)},  \label{FC}
\end{equation}%
where%
\begin{equation}
\sum_{\sigma }=\sum_{s=\pm 1}\sum_{j=\pm 1/2,\ldots }\int_{0}^{\infty
}d\lambda \int_{0}^{\infty }dk.  \label{SumSig}
\end{equation}%
Substituting (\ref{psi-1}) into (\ref{FC}), the expression obtained is
divergent and some regularization procedure is needed. We will assume that a
cutoff function is present without explicitly writing it. The specific form
of this function will not be important for the further discussion.

By making use of the expression (\ref{psi-1}) for the mode functions and the
relation%
\begin{equation}
\sum_{j=\pm 1/2,\ldots }J_{\beta }^{2}(\lambda r)=\sum_{j=\pm 1/2,\ldots
}J_{\beta +\epsilon _{j}}^{2}(\lambda r)=\sum_{j}\left[ J_{qj-1/2}^{2}(%
\lambda r)+J_{qj+1/2}^{2}(\lambda r)\right] ,  \label{SumRel}
\end{equation}%
the FC is presented in the form%
\begin{equation}
\langle 0|\bar{\psi}\psi |0\rangle =\langle 0|\bar{\psi}\psi |0\rangle _{%
\mathrm{s}}+\langle \bar{\psi}\psi \rangle _{\mathrm{b}}.  \label{FCdec}
\end{equation}%
Here the separate terms on the right-hand side are given by the expressions%
\begin{equation}
\langle 0|\bar{\psi}\psi |0\rangle _{\mathrm{s}}=-\frac{qm}{\pi ^{2}}%
\sum_{j}\int_{0}^{\infty }d\lambda \int_{0}^{\infty }dk\frac{\lambda }{%
\omega }[J_{qj-1/2}^{2}(\lambda r)+J_{qj+1/2}^{2}(\lambda r)],  \label{FCs}
\end{equation}%
and%
\begin{eqnarray}
\langle \bar{\psi}\psi \rangle _{\mathrm{b}} &=&\frac{iq}{2\pi ^{2}}%
\sum_{j}\int_{0}^{\infty }d\lambda \int_{0}^{\infty }dk\frac{\lambda }{%
\omega }[J_{qj-1/2}^{2}(\lambda r)+J_{qj+1/2}^{2}(\lambda r)]  \notag \\
&&\times \left[ \left( k-im\right) e^{2ikz}-\left( k+im\right) e^{-2ikz}%
\right] ,  \label{FCb}
\end{eqnarray}%
where and in what follows we identify%
\begin{equation}
\sum_{j}=\sum_{j=1/2,3/2,\ldots }.  \label{jsum}
\end{equation}

First we consider the part $\langle 0|\bar{\psi}\psi |0\rangle _{\mathrm{s}}$%
. This part corresponds to the FC in the boundary-free cosmic string
spacetime (see \cite{Beze08}). The geometry of cosmic string is flat outside
the string core and, hence, the renormalization is reduced to the
substraction of the corresponding quantity in Minkowski spacetime:%
\begin{equation}
\langle \bar{\psi}\psi \rangle _{\mathrm{s,ren}}=\langle 0|\bar{\psi}\psi
|0\rangle _{\mathrm{s}}-\langle 0|\bar{\psi}\psi |0\rangle _{\mathrm{M}}.
\label{FCsren0}
\end{equation}
For the further transformation, we use the representation%
\begin{equation}
\frac{1}{\omega }=\frac{2}{\sqrt{\pi }}\int_{0}^{\infty
}dse^{-(k^{2}+\lambda ^{2}+m^{2})s^{2}}.  \label{rel1}
\end{equation}%
After the substitution of (\ref{rel1}) into (\ref{FCs}) and changing the
order of integrations, the integral over $\lambda $ is expressed in terms of
the modified Bessel function of the first kind, $I_{\nu }(z)$ (see \cite%
{Prud86}). Integrating over $k$ and introducing a new integration variable $%
y=r^{2}/(2s^{2})$, the boundary-free part is presented in the form%
\begin{equation}
\langle 0|\bar{\psi}\psi |0\rangle _{\mathrm{s}}=-\frac{qm}{2\pi ^{2}r^{2}}%
\int_{0}^{\infty }dye^{-y-m^{2}r^{2}/2y}\mathcal{I}(q,y),  \label{FCs0}
\end{equation}%
with the notation
\begin{equation}
\mathcal{I}(q,y)=\sum_{j}\left[ I_{qj-1/2}(y)+I_{qj+1/2}(y)\right] .
\label{Iqy}
\end{equation}

For the series (\ref{Iqy}) we use the formula%
\begin{eqnarray}
\mathcal{I}(q,y) &=&\frac{2}{q}\sideset{}{'}{\sum}_{l=0}^{p}(-1)^{l}\cos
\left( \pi l/q\right) e^{y\cos (2\pi l/q)}  \notag \\
&&+\frac{2}{\pi }\cos \left( q\pi /2\right) \int_{0}^{\infty }dx\frac{\sinh
\left( qx/2\right) \sinh \left( x/2\right) }{\cosh (qx)-\cos (q\pi )}%
e^{-y\cosh x},  \label{SumForm}
\end{eqnarray}%
where $p$ is an integer number defined by $2p\leqslant q<2p+2$ and the prime
on the sign of the sum means that the term $l=0$ should be taken with the
coefficient 1/2. Formula (\ref{SumForm}) is obtained as a special case of
the more general formula derived in \cite{Beze10}. By taking into account
that $\mathcal{I}(1,y)=e^{y}$, we see that the contribution of the term $l=0$
in (\ref{SumForm}) into $\langle 0|\bar{\psi}\psi |0\rangle _{\mathrm{s}}$
coincides with the FC in Minkowski spacetime. The renormalized value of the
FC in the boundary-free cosmic string spacetime is obtained subtracting this
term.

Combining (\ref{FCs0}) and (\ref{SumForm}), after the integration over $y$,
for the renormalized FC in the boundary-free cosmic string spacetime one
finds\footnote{%
An alternative integral representation for $\langle \bar{\psi}\psi \rangle _{%
\mathrm{s,ren}}$ is given in \cite{Beze08}.}
\begin{eqnarray}
&&\langle \bar{\psi}\psi \rangle _{\mathrm{s,ren}}=-\frac{2m^{3}}{\pi ^{2}}%
\bigg[\sum_{l=1}^{p}(-1)^{l}\cos \left( \pi l/q\right) f_{1}(2mr\sin (\pi
l/q))  \notag \\
&&\qquad +\frac{2q}{\pi }\cos \left( q\pi /2\right) \int_{0}^{\infty }dx%
\frac{\sinh \left( qx\right) \sinh (x)f_{1}(2mr\cosh x)}{\cosh (2qx)-\cos
(q\pi )}\bigg],  \label{FCsren}
\end{eqnarray}%
where we have defined the function
\begin{equation}
f_{\nu }(y)=y^{-\nu }K_{\nu }(y).  \label{fnu}
\end{equation}%
Note that for this function one has the relation $\partial _{y}f_{\nu
}(y)=-yf_{\nu +1}(y)$. As it is seen from (\ref{FCsren}), for a massless
field the renormalized FC vanishes in the boundary-free cosmic string
spacetime.

Now we turn to the term $\langle \bar{\psi}\psi \rangle _{\mathrm{b}}$ in (%
\ref{FCdec}) which is induced by the boundary. For points away from the
boundary, this term is finite and the cutoff function, implicitly assumed
before, can be safely removed. For the further transformations it is
convenient to write the boundary-induced term in the FC in the form%
\begin{eqnarray}
\langle \bar{\psi}\psi \rangle _{\mathrm{b}} &=&\frac{q}{2\pi ^{2}}\left(
2m+\partial _{z}\right) \sum_{j}\int_{0}^{\infty }d\lambda \lambda
\int_{0}^{\infty }dk\,\frac{1}{\omega }  \notag \\
&&\times \cos (2kz)[J_{qj-1/2}^{2}(\lambda r)+J_{qj+1/2}^{2}(\lambda r)].
\label{FCb2}
\end{eqnarray}%
Similar to the boundary-free case, by using the representation (\ref{rel1})
and changing the order of integrations, the integrals over $k$ and $\lambda $
are explicitly evaluated. Changing the integration variable to $%
y=r^{2}/(2s^{2})$, we get%
\begin{equation}
\langle \bar{\psi}\psi \rangle _{\mathrm{b}}=\frac{q}{4\pi ^{2}r^{2}}\left(
2m+\partial _{z}\right) \int_{0}^{\infty
}dye^{-(2z^{2}/r^{2}+1)y-m^{2}r^{2}/(2y)}\mathcal{I}(q,y),  \label{FCb3}
\end{equation}%
with the notation (\ref{Iqy}). As it is seen, the boundary-induced part in
the FC does not vanish for a massless field. First we consider this case.

For a massless field the integral in (\ref{FCb3}) is evaluated by using the
formula from \cite{Prud86} and after that the series over $j$ is reduced to
the sum of a geometric progression. In this way we find%
\begin{equation}
\langle \bar{\psi}\psi \rangle _{\mathrm{b}}=-\frac{q}{4\pi ^{2}rz^{2}}\frac{%
u^{q}}{u^{2q}-1}\left( 1+\frac{qz}{\sqrt{r^{2}+z^{2}}}\frac{u^{2q}+1}{%
u^{2q}-1}\right) ,  \label{FCbm0}
\end{equation}%
with the notation%
\begin{equation}
u=z/r+\sqrt{1+z^{2}/r^{2}}.  \label{u}
\end{equation}%
As it is seen, for a massless field the boundary induced part is always
negative. By taking into account that in this case the boundary-free part
vanishes, we conclude that the same is the case for the total FC. For $q=1$,
Eq. (\ref{FCbm0}) is reduced to the known result for the plate in Minkowski
spacetime:%
\begin{equation}
\langle \bar{\psi}\psi \rangle _{\mathrm{b}}^{\mathrm{(M)}}=-\frac{1}{4\pi
^{2}z^{3}}.  \label{FCbm0M}
\end{equation}%
$\langle \bar{\psi}\psi \rangle _{\mathrm{b}}^{\mathrm{(M)}}$ is the leading
term in the asymptotic expansion for $\langle \bar{\psi}\psi \rangle _{%
\mathrm{b}}$ in the limit $z\rightarrow 0$. Assuming $z\ll r$, we have
\begin{equation}
\langle \bar{\psi}\psi \rangle _{\mathrm{b}}\approx -\frac{1}{4\pi ^{2}z^{3}}%
\left[ 1-\frac{(q^{2}-1)(7q^{2}+17)}{360}\left( \frac{z}{r}\right) ^{4}%
\right] .  \label{FCbLarger}
\end{equation}%
For large values of $z$, $z\gg r$, the leading term has the form%
\begin{equation}
\langle \bar{\psi}\psi \rangle _{\mathrm{b}}\approx -\frac{q\left(
1+q\right) }{8\pi ^{2}z^{3}}\left( \frac{r}{2z}\right) ^{q-1}.
\label{FCm0Largz}
\end{equation}%
Note that for $q>1$ the boundary-induced part vanishes on the string.

Now we return to the case of a massive field. In order to provide an
expression of the condensate more convenient for numerical calculations, we
use the formula (\ref{SumForm}). After the substitution into (\ref{FCb3}),
the integral over $y$ is expressed in terms of the Macdonald function. As a
result for the boundary-induced part in the FC we find
\begin{eqnarray}
\langle \bar{\psi}\psi \rangle _{\mathrm{b}} &=&\frac{2m^{3}}{\pi ^{2}}\bigg[%
\sideset{}{'}{\sum}_{l=0}^{p}(-1)^{l}\cos \left( \pi l/q\right)
F(2mz,2mr\sin (\pi l/q))  \notag \\
&&+\frac{2}{\pi }q\cos \left( q\pi /2\right) \int_{0}^{\infty }dx\frac{\sinh
\left( qx\right) \sinh x}{\cosh (2qx)-\cos (q\pi )}F(2mz,2mr\cosh x)\bigg],
\label{FCB4}
\end{eqnarray}%
where%
\begin{equation}
F(x,y)=f_{1}(\sqrt{x^{2}+y^{2}})-xf_{2}(\sqrt{x^{2}+y^{2}}).  \label{Fxy}
\end{equation}%
In this formula, the $l=0$ term corresponds to the FC induced by a boundary
in Minkowski spacetime:%
\begin{equation}
\langle \bar{\psi}\psi \rangle _{\mathrm{b}}^{\mathrm{(M)}}=\frac{m^{2}}{\pi
^{2}}\left( 2m+\partial _{z}\right) f_{1}(2mz)=2\frac{m^{3}}{\pi ^{2}}%
F(2mz,0).  \label{FCMink}
\end{equation}%
Near the boundary this term dominates in (\ref{FCB4}). The remained part
corresponds to the contribution induced by the nontrivial topology of the
string. For $r\neq 0$ the latter is finite on the boundary, $z=0$. In order
to find the asymptotic expression of the boundary-induced FC for points near
the string it is convenient to use the representation (\ref{FCb3}). The
dominant contribution comes from the term with $j=1/2$ in (\ref{Iqy}) and to
the leading order one finds%
\begin{equation}
\langle \bar{\psi}\psi \rangle _{\mathrm{b}}\approx qm^{q+2}r^{q-1}\frac{%
f_{(q+1)/2}(2mz)-2mzf_{(q+3)/2}(2mz)}{2^{(q-1)/2}\pi ^{2}\Gamma
((q+1)/2)}. \label{FCnr0}
\end{equation}%
As it is seen, similar to the case of a massless field, for $q>1$ and $z\neq
0$ the boundary-induced part of the FC vanishes on the string. At large
distances from the boundary, $mz\gg 1$, the boundary-induced part is
exponentially suppressed.

\begin{figure}[tbp]
\begin{center}
\epsfig{figure=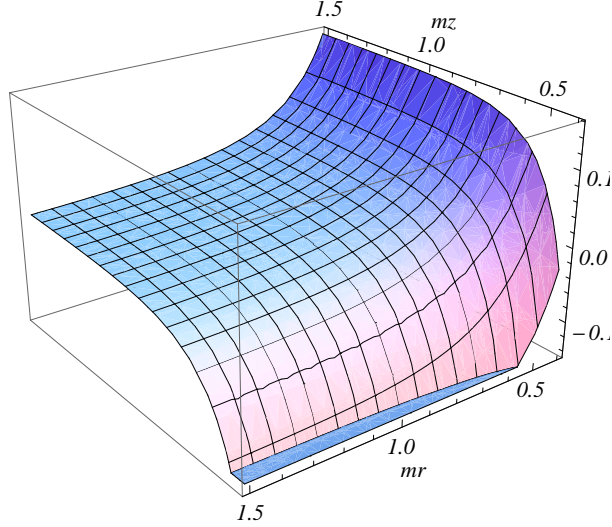,width=8.cm,height=7.cm}
\end{center}
\caption{Renormalized FC, $\langle \bar{\protect\psi}\protect\psi \rangle _{%
\mathrm{ren}}/m^{3}$, in the geometry of a flat boundary in cosmic string
spacetime with $q=3$ as a function of the distances from the boundary and
from the string.}
\label{fig1}
\end{figure}

Having the separate parts, for the renormalized FC we get:%
\begin{equation}
\langle \bar{\psi}\psi \rangle _{\mathrm{ren}}=\langle \bar{\psi}\psi
\rangle _{\mathrm{s,ren}}+\langle \bar{\psi}\psi \rangle _{\mathrm{b}}.
\label{FCren}
\end{equation}
In figure \ref{fig1}, for the cosmic string with $q=3$, we have plotted the
quantity $\langle \bar{\psi}\psi \rangle _{\mathrm{ren}}/m^{3}$ as a
function of distances from the boundary and from the string in units of the
Compton wavelength for the fermionic particle. Near the boundary, the
boundary induced part dominates and the FC is negative. For points near the
string, the FC is dominated by the boundary-free part and it is positive.

\section{Energy-momentum tensor}

\label{sec:EMT}

In this section we consider another important characteristic of the
fermionic vacuum, the VEV of the energy-momentum tensor. This VEV can be
evaluated by using the mode sum formula
\begin{equation}
\left\langle 0\left\vert T_{\mu \nu }\right\vert 0\right\rangle =\frac{i}{2}%
\sum_{\sigma }[\bar{\psi}_{\sigma }^{(-)}\gamma _{(\mu }\nabla _{\nu )}\psi
_{\sigma }^{(-)}-(\nabla _{(\mu }\bar{\psi}_{\sigma }^{(-)})\gamma _{\nu
)}\psi _{\sigma }^{(-)}]\ ,  \label{modesum}
\end{equation}%
where the brackets enclosing the indices mean the simmetrization. Inserting
the expression for the negative-energy mode functions, similar to the case
of the FC, the VEV is presented in the decomposed form:%
\begin{equation}
\left\langle 0\left\vert T_{\mu \nu }\right\vert 0\right\rangle
=\left\langle 0\left\vert T_{\mu \nu }\right\vert 0\right\rangle _{\mathrm{s}%
}+\left\langle T_{\mu \nu }\right\rangle _{\mathrm{b}},  \label{EMTdec}
\end{equation}%
where the first and second terms on the right-hand side correspond to the
boundary-free and boundary-induced parts, respectively. It can be seen that
both parts are diagonal. The latter property is not a consequence of the
problem geometry. As it has been shown in \cite{Beze11}, for a scalar field
obeying Dirichlet and Neumann boundary conditions on a plate at $z=0$, the
off-diagonal component $\left\langle 0\left\vert T_{13}\right\vert
0\right\rangle $ does not vanish for a general curvature coupling parameter $%
\xi $. Only in the special case with $\xi =1/4$ the VEV of the
energy-momentum tensor is diagonal.

\subsection{Boundary-free part}

For the VEV in the boundary-free cosmic string geometry we find the
expression (no summation over $\nu $)%
\begin{equation}
\left\langle 0\left\vert T_{\nu }^{\nu }\right\vert 0\right\rangle _{\mathrm{%
s}}=\frac{q}{\pi ^{2}}\sum_{j}\int_{0}^{\infty }dk\,\int_{0}^{\infty
}d\lambda \frac{\lambda ^{3}}{\omega }g_{\beta }^{(\nu )}(\lambda r),
\label{EMTs}
\end{equation}%
with the notations%
\begin{eqnarray}
g_{\beta }^{(0)}(y) &=&\frac{\omega ^{2}}{k^{2}}g_{\beta }^{(3)}(y)=-\frac{%
\omega ^{2}}{\lambda ^{2}}[J_{\beta }^{2}(y)+J_{\beta +1}^{2}(y)],  \notag \\
g_{\beta }^{(1)}(y) &=&J_{\beta }^{2}(y)+J_{\beta +1}^{2}(y)-\frac{2qj}{y}%
J_{\beta }(y)J_{\beta +1}(y),  \label{genu} \\
g_{\beta }^{(2)}(y) &=&\frac{2qj}{y}J_{\beta }(y)J_{\beta +1}(y).  \notag
\end{eqnarray}%
For $r\neq 0$ the renormalization of the boundary-free part is reduced to
the subtraction of the corresponding VEV in Minkowski spacetime:%
\begin{equation}
\left\langle T_{\mu }^{\nu }\right\rangle _{\mathrm{s,ren}}=\left\langle
0\left\vert T_{\mu }^{\nu }\right\vert 0\right\rangle _{\mathrm{s}%
}-\left\langle 0\left\vert T_{\mu }^{\nu }\right\vert 0\right\rangle _{%
\mathrm{M}}.  \label{EMTsren}
\end{equation}%
In Appendix \ref{sec:Appendix} we show that the renormalized boundary-free
VEV is presented in the form (no summation over $\nu $)%
\begin{eqnarray}
\left\langle T_{\nu }^{\nu }\right\rangle _{\mathrm{s,ren}} &=&\frac{2m^{4}}{%
\pi ^{2}}\bigg[\sum_{l=1}^{p}(-1)^{l}\cos \left( \pi l/q\right) F_{\nu
}^{(0)}(2mr\sin (\pi l/q))  \notag \\
&&+\frac{2q}{\pi }\cos \left( q\pi /2\right) \int_{0}^{\infty }dx\frac{\sinh
\left( qx\right) \sinh (x)F_{\nu }^{(0)}(2mr\cosh x)}{\cosh (2qx)-\cos (q\pi
)}\bigg],  \label{EMTsren1}
\end{eqnarray}%
where we have defined%
\begin{eqnarray}
F_{0}^{(0)}(y) &=&F_{1}^{(0)}(y)=F_{3}^{(0)}(y)=f_{2}(y),  \notag \\
F_{2}^{(0)}(y) &=&-3f_{2}(y)-f_{1}(y),  \label{F20}
\end{eqnarray}%
with the notation (\ref{fnu}). For $q<2$ the first term in the square
brackets is absent and the formula (\ref{EMTsren1}) is reduced to the one
derived in \cite{Beze06}. For general case of $q$ an alternative integral
representation is given in \cite{Beze08}. For a massless field the
corresponding renormalized VEV for the energy-momentum tensor was found in
\cite{Frol87,Dowk87b}:
\begin{equation}
\left\langle T_{\mu }^{\nu }\right\rangle _{\mathrm{s,ren}}=-\frac{%
(q^{2}-1)(7q^{2}+17)}{2880\pi ^{2}r^{4}}\mathrm{diag}(1,1,-3,1).
\label{T00srenm0}
\end{equation}%
Fermionic propagators for a massive field are considered in Refs. \cite%
{Line95,More95,Beze06}. Note that in the boundary-free cosmic string
spacetime one has the relation $\left\langle T_{0}^{0}\right\rangle _{%
\mathrm{s,ren}}=\left\langle T_{3}^{3}\right\rangle _{\mathrm{s,ren}}$ for a
massive field. The latter is a consequence of the boost invariance along the
axis of the cosmic string.

For a massive field and for points near the string, $mr\ll 1$, the leading
term in the corresponding asymptotic expansion is given by (\ref{T00srenm0})
and the renormalized VEV diverges on the string as $r^{-4}$. At large
distances, $mr\gg 1$, for $q<2$ the first term in the square brackets in (%
\ref{EMTsren1}) is absent and the dominant contribution in the second term
comes from the region near the lower limit of the integration. In this case $%
\left\langle T_{\mu }^{\nu }\right\rangle _{\mathrm{s,ren}}$ is suppressed
by the factor $e^{-2mr}$. For $q>2$ the dominant contribution at large
distances comes from the $l=1$ term in (\ref{EMTsren1}) and the suppression
factor is $e^{-2mr\sin (\pi /q)}$. Note that the contribution of the first
term in the square brackets of (\ref{EMTsren1}) is always negative for $\nu
=0,1,3$, and positive for $\nu =2$. The contribution from the second term
has opposite sign for $3+4n<q<5+4n$ with $n=0,1,2,\ldots $. However, for
large $q$ this contribution is small.

\begin{figure}[tbp]
\begin{center}
\epsfig{figure=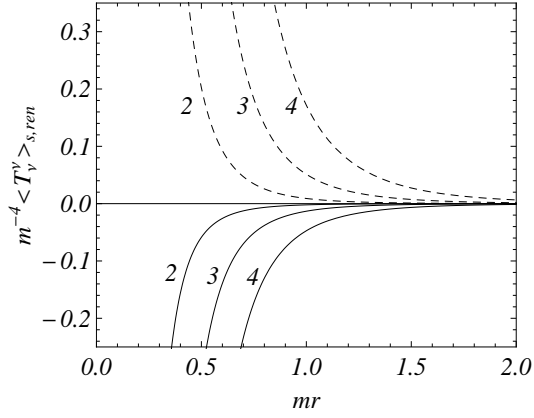,width=7.cm,height=5.5cm}
\end{center}
\caption{Boundary-free parts in the energy density (full curves) and the
azimuthal stress (dashed curves) as functions of the distance from the
string for separate values of the parameter $q$ (numbers near the curves).}
\label{fig2}
\end{figure}

In figure \ref{fig2} we plotted the boundary-free parts in the
energy density (full curves) and the azimuthal stress (dashed
curves) as functions of the distance from the string in units of
the Compton wavelength. The numbers near the curves correspond to
the values of the parameter $q$. Note that the quantity (no
summation over $\nu $) $-\left\langle T_{\nu }^{\nu }\right\rangle
_{\mathrm{s,ren}}$, with $\nu =1,2,3$, is the vacuum effective
pressure along the corresponding direction. Hence, in the
boundary-free cosmic string spacetime the vacuum azimuthal
pressure is negative, whereas pressures along the radial and axial
directions are positive.

\subsection{Boundary-induced part}

For the boundary-induced part of the vacuum energy-momentum tensor we have
(no summation over $\nu =0,1,2$)%
\begin{equation}
\left\langle T_{\nu }^{\nu }\right\rangle _{\mathrm{b}}=-\frac{q}{\pi ^{2}}%
m\left( m+\partial _{2z}\right) \sum_{j}\int_{0}^{\infty }d\lambda
\int_{0}^{\infty }dk\frac{\lambda ^{3}}{\omega }g_{\beta }^{(\nu )}(\lambda
r)\frac{\cos (2kz)}{k^{2}+m^{2}},  \label{EMTb}
\end{equation}%
with the functions $g_{\beta }^{(\nu )}(y)$ defined in
(\ref{genu}), and the corresponding axial stress vanishes:
\begin{equation}
\left\langle T_{3}^{3}\right\rangle _{\mathrm{b}}=0.  \label{T33b}
\end{equation}%
For a massless field the boundary-induced part in the VEV of the
energy-momentum tensor vanishes. Note that the latter is not the case for a
massless scalar field (see \cite{Beze11}). It can be explicitly checked that
the VEV (\ref{EMTb}) obeys the covariant conservation equation $\nabla _{\nu
}\langle T_{\mu }^{\nu }\rangle _{\mathrm{b}}=0$, which for the problem
under consideration is reduced to the single equation%
\begin{equation}
\partial _{r}(r\left\langle T_{1}^{1}\right\rangle _{\mathrm{b}%
})=\left\langle T_{2}^{2}\right\rangle _{\mathrm{b}}.  \label{ContEq}
\end{equation}%
We have also the trace relation%
\begin{equation}
\left\langle T_{\nu }^{\nu }\right\rangle _{\mathrm{b}}=m\langle \bar{\psi}%
\psi \rangle _{\mathrm{b}}.  \label{EMTtrace}
\end{equation}

Further transformations for the components of the boundary-induced part in
the VEV of the energy-momentum tensor are given in Appendix \ref%
{sec:Appendix}. They are presented in the form (no summation over $\nu $)%
\begin{eqnarray}
\left\langle T_{\nu }^{\nu }\right\rangle _{\mathrm{b}} &=&\frac{2m^{4}}{\pi
^{2}}\bigg[\sideset{}{'}{\sum}_{l=0}^{p}(-1)^{l}\cos \left( \pi l/q\right)
F_{\nu }(2mz,2mr\sin (\pi l/q))  \notag \\
&&+\frac{2q}{\pi }\cos \left( q\pi /2\right) \int_{0}^{\infty }dx\frac{\sinh
\left( qx\right) \sinh \left( x\right) }{\cosh (2qx)-\cos (q\pi )}F_{\nu
}(2mz,2mr\cosh x)\bigg],  \label{EMTb1}
\end{eqnarray}%
where%
\begin{eqnarray}
F_{0}(x,y) &=&F_{1}(x,y)=-G_{2}(x,y),  \notag \\
F_{2}(x,y) &=&2G_{2}(x,y)+F(x,y),  \label{F01} \\
\;F_{3}(x,y) &=&0.  \notag
\end{eqnarray}%
In (\ref{EMTb1}), the function $G_{\nu }(x,y)$ is defined as%
\begin{equation}
G_{\nu }(x,y)=e^{x}\int_{x}^{\infty }du\,e^{-u}f_{\nu }(\sqrt{u^{2}+y^{2}}),
\label{Genu}
\end{equation}%
and the function $F(x,y)$ is given by (\ref{Fxy}). As we see, the
boundary-induced part in the energy density is equal to the radial stress: $%
\left\langle T_{0}^{0}\right\rangle _{\mathrm{b}}=\left\langle
T_{1}^{1}\right\rangle _{\mathrm{b}}$ and the axial stress vanishes. By
using the relation (\ref{GenuRel}) we can see that $\partial
_{y}[yF_{1}(x,y)]=$ $F_{2}(x,y)$. With this relation the covariant
conservation equation (\ref{ContEq}) is explicitly checked.

Alternative expressions for the boundary-induced parts in the energy density
and in the azimuthal stress are given by formulas (\ref{T00b2}) and (\ref%
{T22b3}) in Appendix \ref{sec:Appendix}. In particular, from (\ref{T00b2})
it follows that the boundary-induced part in the energy density is always
negative for $r>0$. For the investigation of the boundary-induced VEV near
the cosmic string it is more convenient to use the representations (\ref%
{T00b2}) and (\ref{T22b3}). The leading contributions come from the terms
with $j=1/2$ in the series for $\mathcal{I}(q,y)$. To the leading order one
gets%
\begin{equation}
\left\langle T_{\mu }^{\nu }\right\rangle _{\mathrm{b}}\approx \frac{%
qm^{q+3}r^{q-1}}{2^{(q-1)/2}\pi ^{2}}\frac{%
f_{(q+1)/2}(2mz)-2mzf_{(q+3)/2}(2mz)}{\left( q+2\right) \Gamma ((q+1)/2)}%
\mathrm{diag}(1,1,q,0),  \label{EMTr0}
\end{equation}
and the boundary-induced part vanishes on the string for $q>1$ and $z\neq 0$%
. Note that from the relation (\ref{Genux0}) in Appendix \ref{sec:Appendix}
it follows that the function in the numerator of (\ref{EMTr0}) is negative.

For $q=1$ there is only contribution from the term with $l=0$ and from (\ref%
{EMTb1}) we obtain the result for a plate in Minkowski spacetime (note that
in this case the asymptotic expression (\ref{EMTr0}) is exact):%
\begin{equation}
\left\langle T_{\mu }^{\nu }\right\rangle _{\mathrm{b}}^{\mathrm{(M)}}=\frac{%
m^{4}}{3\pi ^{2}}[f_{1}(2mz)-2mzf_{2}(2mz)]\,\mathrm{diag}(1,1,1,0).
\label{EMTM}
\end{equation}%
In this case the vacuum stresses along the directions parallel to the plate
are equal to the the energy density. This result is a consequence of the
Lorentz invariance of the problem along these directions. Note that in the
presence of the cosmic string this invariance is broken. However, the radial
stress remains equal to the energy density. Formula (\ref{EMTM}) is a
special case of a more general result given in \cite{Eliz11} for a plate in
flat spacetime with toroidally compactified spatial dimensions.

Extracting the term $l=0$ in (\ref{EMTb1}), the boundary-induced part in the
VEV of the energy-momentum tensor can be written in the form%
\begin{equation}
\left\langle T_{\mu }^{\nu }\right\rangle _{\mathrm{b}}=\left\langle T_{\mu
}^{\nu }\right\rangle _{\mathrm{b}}^{\mathrm{(M)}}+\left\langle T_{\mu
}^{\nu }\right\rangle _{\mathrm{b}}^{\mathrm{(s)}},  \label{EMTbdec}
\end{equation}%
where $\left\langle T_{\mu }^{\nu }\right\rangle _{\mathrm{b}}^{\mathrm{(s)}%
} $ comes from the nontrivial topology of the cosmic string spacetime. We
can see that the latter is finite everywhere except at the point $r=z=0$.
The divergences on the boundary are contained in the Minkowskian part. Near
the boundary, for $z\ll r$, to the leading order one has (no summation over $%
\nu =0,1,2$):%
\begin{equation}
\left\langle T_{\nu }^{\nu }\right\rangle _{\mathrm{b}}\approx \left\langle
T_{\nu }^{\nu }\right\rangle _{\mathrm{b}}^{\mathrm{(M)}}\approx -\frac{m}{%
12\pi ^{2}z^{3}}.  \label{EMTz0}
\end{equation}

\begin{figure}[tbph]
\begin{center}
\begin{tabular}{cc}
\epsfig{figure=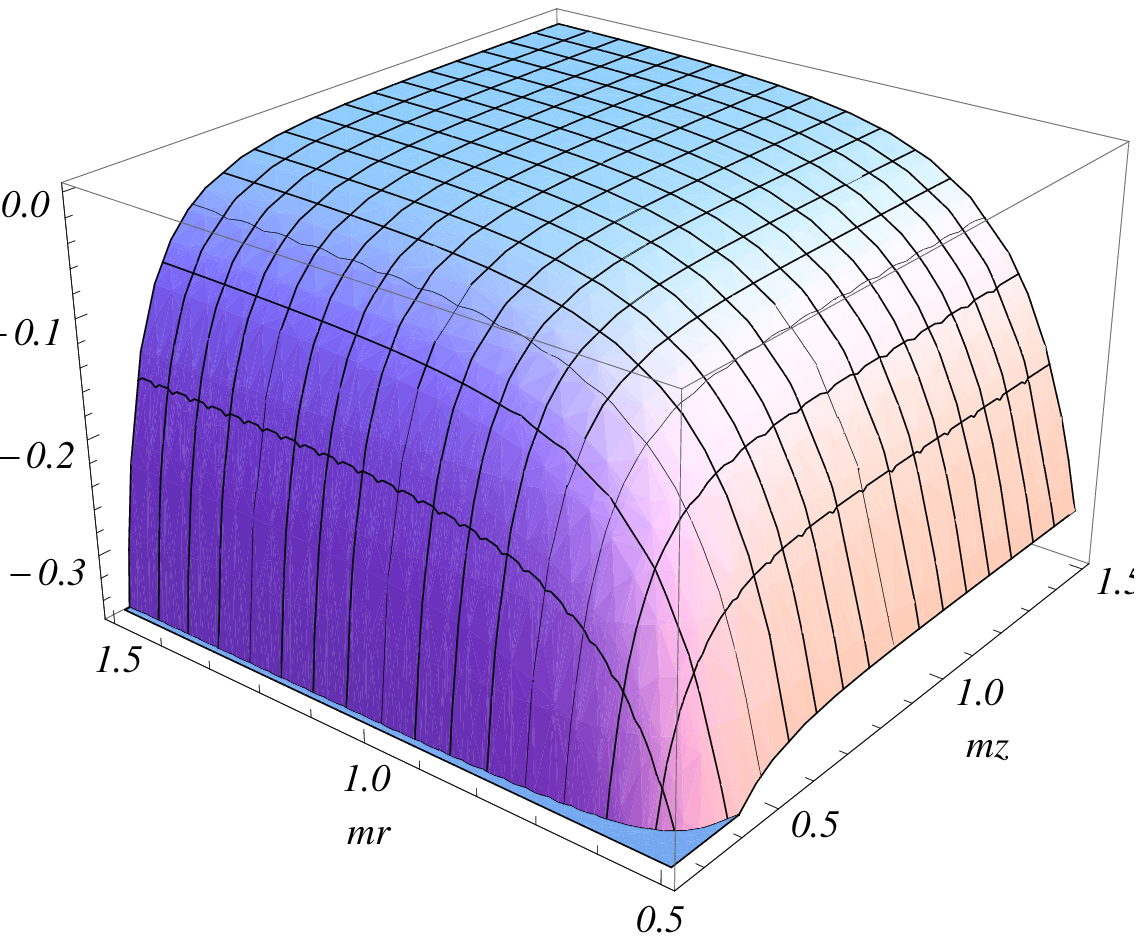,width=7.cm,height=6.cm} & \quad %
\epsfig{figure=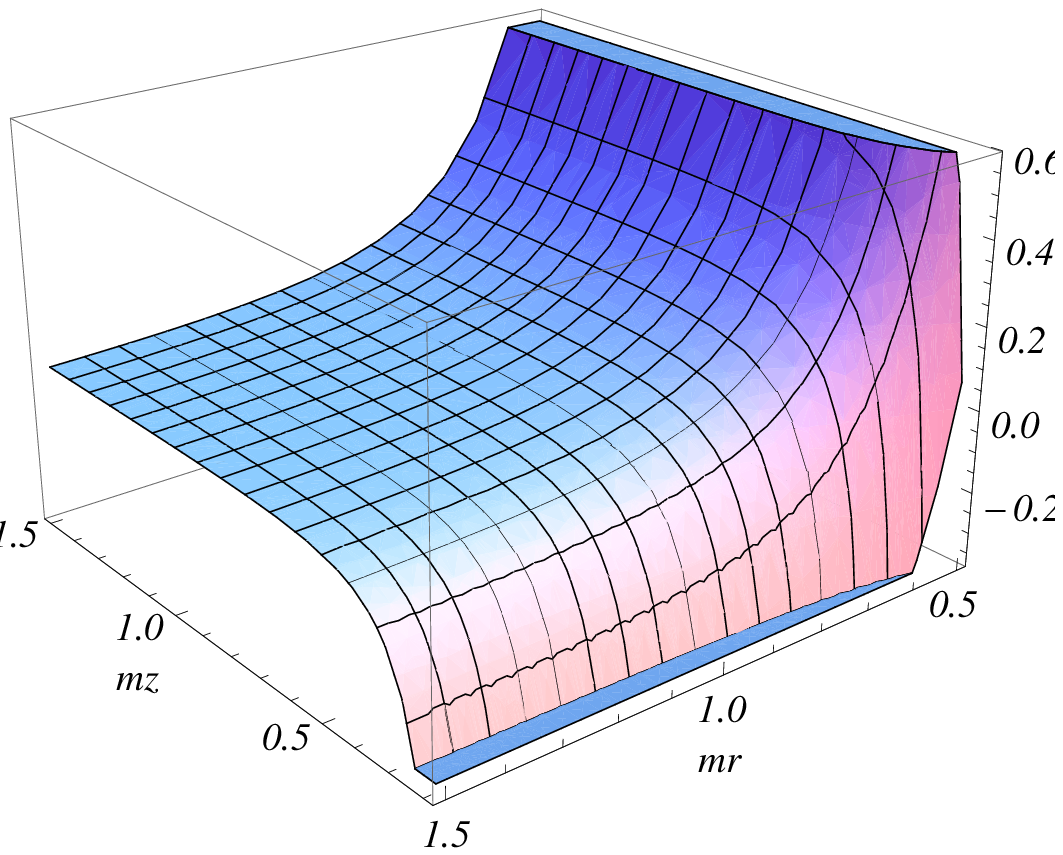,width=7.cm,height=6.cm}%
\end{tabular}%
\end{center}
\caption{VEV of the the energy density, $m^{-4}\langle T_{0 }^{0 }\rangle _{%
\mathrm{ren}}$ (left plot), and the azimuthal stress, $m^{-4}\langle T_{2
}^{2 }\rangle _{\mathrm{ren}}$ (right plot), as functions of the distances
from the string and from the boundary for $q=3$.}
\label{fig3}
\end{figure}

For points away from the boundary the renormalized VEV of the
energy-momentum tensor is presented as
\begin{equation}
\left\langle T_{\mu }^{\nu }\right\rangle _{\mathrm{ren}}=\left\langle
T_{\mu }^{\nu }\right\rangle _{\mathrm{s,ren}}+\left\langle T_{\mu }^{\nu
}\right\rangle _{\mathrm{b}},  \label{EMTren}
\end{equation}%
where the boundary-free and boundary-induced parts are given by (\ref%
{EMTsren1}) and (\ref{EMTb1}), respectively. The dependence of the
corresponding energy density (left plot) and the azimuthal stress
(right plot) on the distances from the string and from the
boundary is presented in figure \ref{fig3} for the geometry of
cosmic string with $q=3$. In the case of the energy density both
the boundary-free and boundary-induced parts are negative. For the
azimuthal stress, the boundary-free part is positive whereas the
boundary-induced part is negative. The latter dominates for points
near the boundary.

\section{Conclusion}

\label{sec:Conc}

In this paper we have considered the interplay of topological and
boundary-induced quantum effects in the fermionic vacuum. The
nontrivial topology of the background spacetime is due to the
presence of an idealized cosmic string and the fermionic field
obeys MIT bag boundary condition on a plane perpendicular to the
string. As physical characteristics of the vacuum state, we have
investigated the FC and the VEV\ of the energy-momentum tensor.
For the evaluation of these VEVs a complete set of mode functions
for the corresponding boundary value problem is necessary. A
crucial point in this paper was the obtainment of both positive-
and negative-energy wave functions in section \ref{sec:Modes}.
These functions are given by the expressions (\ref{psi+1}) and
(\ref{psi-1}), respectively. They can be used for the
investigation of various quantum electrodynamics processes around
the cosmic string in the presence of boundary involving electrons
and positrons.

The FC is decomposed into the boundary-free and boundary-induced
contributions. The renormalized boundary-free part is given by (\ref{FCsren}%
) and, for a general value of $q$, it vanishes for a massless field. As to
the boundary-induced part, it is finite for points outside of boundary and
is given by the formula (\ref{FCB4}). An alternative representation is
provided by (\ref{FCb3}). The expression for the boundary-induced part in FC
is simplified for the case of a massless field, Eq. (\ref{FCbm0}). In this
case the boundary-free part vanishes and the boundary-induced part is
negative for points outside the string. For points near the string the
boundary-induced part in FC behaves as $r^{q-1}$. The latter is the case for
a massive field also. On the boundary the FC is divergent. For points near
the boundary the part $\langle \bar{\psi}\psi \rangle _{\mathrm{b}}^{\mathrm{%
(M)}}$ dominates and the FC is negative. The remained part in $\langle \bar{%
\psi}\psi \rangle _{\mathrm{b}}$ corresponds to the contribution induced by
the nontrivial topology of the string and for $r\neq 0$ it is finite on the
boundary. At large distance from the boundary and for a massless field the
leading term in the boundary-induced FC is given by (\ref{FCm0Largz}).
Compared with the case of the Minkowski bulk, the presence of the cosmic
string leads to an additional suppression in the form of the factor $%
(r/2z)^{q-1}$. For a massive field and at large distances from the boundary,
the boundary-induced part of FC is exponentially suppressed.

Another calculation developed in this paper is the VEV of the
energy-momentum tensor. Similar to the case of FC we have the decomposition (%
\ref{EMTdec}). In this way, for points away from the boundary, the
renormalization is reduced to the one for the boundary-free part. For the
general case of the parameter $q$, the renormalized value for the latter is
given by (\ref{EMTsren1}). This formula generalizes various special cases,
previously discussed in the literature. For points near the string the pure
string part diverges as $r^{-4}$. In the opposite limit, at distances larger
than the Compton wavelength for a fermionic particle, $\left\langle T_{\mu
}^{\nu }\right\rangle _{\mathrm{s,ren}}$ is suppressed by the factor $%
e^{-2mr}$ for $q<2$ and by the factor $e^{-2mr\sin (\pi /q)}$ for $q>2$.

The boundary-induced part in the VEV of the energy-momentum tensor, $%
\left\langle T_{\mu }^{\nu }\right\rangle _{\mathrm{b}}$, is given
by the expression (\ref{EMTb1}). This part is diagonal and
vanishes for a massless field. Note that for a scalar field with
general curvature coupling parameter the part in the VEV of the
energy-momentum tensor induced by Dirichlet and Neumann boundaries
perpendicular to the plate has nonzero off-diagonal component
$\left\langle T_{3}^{1}\right\rangle _{\mathrm{b}}$ (see
\cite{Beze11}). The latter is zero in the special case of the
curvature coupling parameter $\xi =1/4$. Another difference,
compared with the fermionic case, is that for a massless scalar
field the boundary-induced part does not vanish. For a massive
fermionic field the boundary-induced part in the radial stress is
equal to the energy density and the axial stress vanishes. For
$q>1$ and $z\neq 0$, the boundary-induced part vanishes on the
string with the asymptotic behavior given by (\ref{EMTr0}). Near
the boundary the VEV is dominated by the part corresponding to a
plate in Minkowski spacetime. The latter is given by the
expression (\ref{EMTM}). The corresponding radial and azimuthal
stresses are equal to the energy density and the latter is
negative. For points near the string, the boundary-free part in
the vacuum energy-momentum tensor dominates. The corresponding
energy density is negative, whereas the azimuthal stress is
positive.

\section*{Acknowledgments}

E.R.B.M. thanks Conselho Nacional de Desenvolvimento Cient\'{\i}fico e Tecnol%
\'{o}gico (CNPq) for partial financial support. A.A.S. was supported by CNPq.

\appendix

\section{Evaluation of the energy density and azimuthal stress}

\label{sec:Appendix}

In this appendix we give some details related with the
transformation of the expressions for the boundary-induced parts
in the VEVs of the energy density and the azimuthal stress. First
we consider the energy density. Our starting point is the
expression (\ref{EMTb}) with $g_{\beta }^{(0)}(\lambda ,\lambda
r)$ defined in (\ref{genu}). As the first step we insert the
representation%
\begin{equation}
\omega =-\frac{2}{\sqrt{\pi }}\int_{0}^{\infty }ds\,\partial
_{s^{2}}e^{-\left( k^{2}+\lambda ^{2}+m^{2}\right) s^{2}}.  \label{omRep}
\end{equation}%
Changing the order of integrations in (\ref{EMTb}), the integral over $%
\lambda $ is expressed in terms of the modified Bessel function of the first
kind. After the integration by parts in the integral over $s$, we find the
expression%
\begin{eqnarray}
\left\langle T_{0}^{0}\right\rangle _{\mathrm{b}} &=&-\frac{qm}{\sqrt{2\pi }%
\pi ^{2}r^{3}}\left( m+\partial _{2z}\right) \int_{0}^{\infty }dy\,\sqrt{y}%
e^{-y}  \notag \\
&&\times \mathcal{I}(q,y)\int_{0}^{\infty }dk\,\frac{\cos (2kz)}{k^{2}+m^{2}}%
e^{-\left( k^{2}+m^{2}\right) r^{2}/2y},  \label{T00b1}
\end{eqnarray}%
with $\mathcal{I}(q,y)$ defined in (\ref{Iqy}). The presence of the factor $%
(k^{2}+m^{2})^{-1}$ in the integrand of (\ref{T00b1}) leads to some
complication in the further evaluation of the VEV when compared with the
case of the FC.

Here we use the relation%
\begin{equation}
\left( m+\partial _{x}\right) \frac{\cos (kx)}{k^{2}+m^{2}}%
=e^{mx}\int_{x}^{\infty }dt\,e^{-mt}\cos (kt).  \label{Rel2}
\end{equation}%
Substituting this into (\ref{T00b1}) we first integrate over $k$ with the
result:%
\begin{equation}
\left\langle T_{0}^{0}\right\rangle _{\mathrm{b}}=-\frac{qm}{\pi ^{2}r^{4}}%
e^{2mz}\int_{z}^{\infty }dte^{-2mt}\int_{0}^{\infty
}dy\,ye^{-m^{2}r^{2}/(2y)-y(2t^{2}/r^{2}+1)}\mathcal{I}(q,y).  \label{T00b2}
\end{equation}%
Now, by using the formula (\ref{SumForm}), the integral over $y$ is given in
terms of the Macdonald function and we come to the final expression
\begin{eqnarray}
\left\langle T_{0}^{0}\right\rangle _{\mathrm{b}} &=&-\frac{2m^{4}}{\pi ^{2}}%
\bigg[\sideset{}{'}{\sum}_{l=0}^{p}(-1)^{l}\cos \left( \pi l/q\right)
G_{2}(2mz,2mr\sin (\pi l/q))  \notag \\
&&+\frac{2q}{\pi }\cos \left( q\pi /2\right) \int_{0}^{\infty }dx\frac{\sinh
\left( qx\right) \sinh \left( x\right) }{\cosh (2qx)-\cos (q\pi )}%
G_{2}(2mz,2mr\cosh x)\bigg],  \label{T00b3}
\end{eqnarray}%
where the function $G_{\nu }(x,y)$ is defined by (\ref{Genu}). By using the
relations%
\begin{eqnarray}
\partial _{x}f_{\nu }(x) &=&-xf_{\nu +1}(x),  \notag \\
x\partial _{x}f_{\nu }(x) &=&-f_{\nu -1}(x)-2\nu f_{\nu }(x),  \label{fnurel}
\end{eqnarray}%
for this function $G_{\nu }(x,y)$ we get the following formula%
\begin{equation}
y^{2}G_{\nu +1}(x,y)=\left( 2\nu -1\right) G_{\nu }(x,y)+f_{\nu -1}(\sqrt{%
x^{2}+y^{2}})-xf_{\nu }(\sqrt{x^{2}+y^{2}}).  \label{GenuRel}
\end{equation}%
In particular, $G_{\nu }(x,0)$ is expressed in terms of the Macdonald
function (see also \cite{Prud86}):%
\begin{equation}
G_{\nu }(x,0)=\frac{xf_{\nu }(x)-f_{\nu -1}(x)}{2\nu -1}.  \label{Genux0}
\end{equation}

Now we turn to the azimuthal stress. The corresponding expression from (\ref%
{EMTb}), with $g_{\beta }^{(2)}(\lambda ,y)$ defined in (\ref{genu}), can be
written in the form
\begin{eqnarray}
\left\langle T_{22}\right\rangle _{\mathrm{b}} &=&2m\frac{q^{2}}{\pi ^{2}}%
\left( m+\partial _{2z}\right) \sum_{j}j\left( 2qj-1-r\partial _{r}\right)
\notag \\
&&\times \int_{0}^{\infty }dk\,\frac{\cos (2kz)}{k^{2}+m^{2}}%
\int_{0}^{\infty }d\lambda \frac{\lambda }{\omega }J_{qj-1/2}^{2}(\lambda r).
\label{T22b1}
\end{eqnarray}%
By using the representation (\ref{rel1}) we evaluate the integral over $%
\lambda $ in terms of the modified Bessel function. Making use of the
relations%
\begin{equation}
(qj-1/2-y\partial
_{y})e^{-y}I_{qj-1/2}(y)=ye^{-y}[I_{qj-1/2}(y)-I_{qj+1/2}(y)],
\label{relBes}
\end{equation}%
and
\begin{equation}
qj[I_{qj-1/2}(y)-I_{qj+1/2}(y)]=(y\partial
_{y}-y+1/2)[I_{qj-1/2}(y)+I_{qj+1/2}(y)],  \label{relBes1}
\end{equation}%
we arrive to the expression%
\begin{eqnarray}
\left\langle T_{2}^{2}\right\rangle _{\mathrm{b}} &=&-\frac{4mq}{\pi ^{5/2}}%
\int_{0}^{\infty }ds\,y^{2}e^{-y}(y\partial _{y}-y+1/2)\mathcal{I}(q,y)
\notag \\
&&\times \int_{0}^{\infty }dk\,\left( m+\partial _{2z}\right) \frac{\cos
(2kz)}{k^{2}+m^{2}}e^{-(k^{2}+m^{2})s^{2}},  \label{T22b2}
\end{eqnarray}%
where $y=r^{2}/(2s^{2})$. The $k$-integration is explicitly done by using
the representation (\ref{Rel2}) with the result%
\begin{eqnarray}
\left\langle T_{2}^{2}\right\rangle _{\mathrm{b}} &=&-2\frac{qm}{\pi ^{2}}%
e^{2mz}\int_{z}^{\infty }dte^{-2mt}\int_{0}^{\infty }dy\,y  \notag \\
&&\times e^{-m^{2}r^{2}/2y-y(2t^{2}/r^{2}+1)}\left( y\partial
_{y}-y+1/2\right) \mathcal{I}(q,y).  \label{T22b3}
\end{eqnarray}%
Now we employ the formula (\ref{SumForm}). After the integration over $y$ we
find the expression%
\begin{eqnarray}
\left\langle T_{2}^{2}\right\rangle _{\mathrm{b}} &=&\frac{2m^{4}}{\pi ^{2}}%
\bigg[\sideset{}{'}{\sum}_{l=0}^{p}(-1)^{l}\cos \left( \pi l/q\right)
F_{2}(2mz,2mr\sin \left( \pi l/q\right) )  \notag \\
&&+\frac{2q}{\pi }\cos \left( q\pi /2\right) \int_{0}^{\infty }dx\frac{\sinh
\left( qx\right) \sinh x}{\cosh (2qx)-\cos (q\pi )}F_{2}(2mz,2mr\cosh x)%
\bigg],  \label{T22b4}
\end{eqnarray}%
with the notation%
\begin{equation}
F_{2}(x,y)=y^{2}G_{3}(x,y)-G_{2}(x,y).  \label{F2}
\end{equation}%
An alternative expression (\ref{F01}) for the function $F_{2}(x,y)$ is
obtained by using the relation (\ref{GenuRel}).

Having the energy density and the azimuthal stress we can find the radial
stress with the help of the trace relation (\ref{EMTtrace}). By using the
expression (\ref{FCB4}) for the FC, we can see that the boundary-induced
part in the radial stress is equal to the energy density.

Note that the transformations described above can also be done for the
boundary-free parts in the energy density and the azimuthal stress. By using
(\ref{rel1}) we find%
\begin{eqnarray}
\left\langle 0|T_{0}^{0}|0\right\rangle _{\mathrm{s}} &=&\frac{q}{2\pi
^{2}r^{4}}\int_{0}^{\infty }dy\,ye^{-m^{2}r^{2}/2y-y}\mathcal{I}(q,y),
\notag \\
\left\langle 0|T_{2}^{2}|0\right\rangle _{\mathrm{s}} &=&\frac{q}{\pi
^{2}r^{4}}\int_{0}^{\infty }dy\,ye^{-m^{2}r^{2}/2y-y}\,(y\partial _{y}-y+1/2)%
\mathcal{I}(q,y).  \label{T22s}
\end{eqnarray}%
Now substituting (\ref{SumForm}), we see that the contribution of the $l=0$
term corresponds to the VEV in boundary-free Minkowski spacetime. The latter
is subtracted in the renormalization procedure and for the renormalized VEVs
of the energy density and the azimuthal stress we find the representation (%
\ref{EMTsren1}). From the boost invariance along the axis of the cosmic
string we have $\left\langle 0|T_{3}^{3}|0\right\rangle _{\mathrm{s,ren}%
}=\left\langle 0|T_{0}^{0}|0\right\rangle _{\mathrm{s,ren}}$. The radial
stress is found from the trace relation $\left\langle 0|T_{\nu }^{\nu
}|0\right\rangle _{\mathrm{s,ren}}=m\langle \bar{\psi}\psi \rangle _{\mathrm{%
s,ren}}$, by using the expression (\ref{FCsren}) for the FC.

\end{document}